\documentclass[aps,prd,prabib,twocolumn,showpacs,nofootinbib]{revtex4}
\usepackage{graphicx} \usepackage{amsmath} \usepackage{amssymb}
\usepackage{amsfonts} \usepackage{bm}

\begin{document}

\newcommand{\be}{\begin{equation}} \newcommand{\ee}{\end{equation}}
\newcommand{\bea}{\begin{eqnarray}}\newcommand{\eea}{\end{eqnarray}}

\title{Non-commutativity as a measure of inequivalent quantization}

\author{Pulak Ranjan Giri} \email{pulakranjan.giri@saha.ac.in}

\affiliation{Theory Division, Saha Institute of Nuclear Physics,
1/AF Bidhannagar, Calcutta 700064, India}

\begin{abstract}
We show that the strength of  non-commutativity could play a role in
determining the boundary condition of a physical problem. As a toy
model we consider the inverse square problem  in non-commutative
space. The scale invariance of the system is known to be explicitly
broken by the scale of non-commutativity $\Theta$. The resulting
problem in non-commutative space is analyzed. It is shown that
despite the presence of  higher singular potential coming from the
leading term of the expansion of the potential to first order in
$\Theta$, it can have a self-adjoint extensions. The boundary
conditions are obtained, belong to a $1$-parameter family and
related to the strength of non-commutativity.
\end{abstract}



\date{\today}

\maketitle


Study of non-commutative spacetime \cite{michael,calmet}  is a
fascinating subject. The expectation that the spacetime could be
non-commutative at small length scale has further accelerated the
research work in this direction. Due to the non-commutativity of
coordinates on a plane $(x,y)$ there exists an uncertainty relation
\begin{eqnarray}
\Delta x\Delta y\sim \Theta\,,
\end{eqnarray}
where $\Theta$ is the non-commutativity parameter. Non-commutativity
to a charged particle can arise due to the nontrivial nature of
spacetime at small length scale or it may arise if the magnetic
field, subjected perpendicular to the plane, is strong enough.
However the idea of non-commutativity of spacetime is quite old way
back in 1947 \cite{snyder}, although that did not get much attention
then. In quantum theory non-commutativity is a key object, for
example coordinate $x$ and its conjugate $p$ are non-commutative,
\begin{eqnarray}
\Delta x\Delta p\sim \hbar\,.
\end{eqnarray}
Even the generalized momenta $P_i$ in the magnetic field, $B$,
background do not commute
\begin{eqnarray}
\Delta P_1\Delta P_2\sim B\,.
\end{eqnarray}
The coordinates of a plane behave as canonical conjugate pairs and
therefore do not commute  in presence of a strong magnetic field
perpendicular to the plane.

The strength of non-commutativity, $\Theta$, may have an intrinsic
origin in spacetime or it may have origin in external magnetic field
as stated before. However, the length scale, $\Theta$, introduced in
the problem due to the non-commutativity  can be exploited to heal
the ultraviolet divergence of the problem under study. In a recent
paper \cite{pulak} we investigated the inverse square problem,
$H=\boldsymbol{p}^2 +\alpha\boldsymbol{r}^{-2}$, in non-commutative
space in order to show how the length scale $\Theta$ can be
successfully used to regularize the problem. Since the inverse
square problem does not possess any dimensional parameter to start
with it is a scale invariant problem. It can be understood from the
transformation $\bf{r}\to\varepsilon\bf{r}$ and
$t\to\varepsilon^2t$. The parameter $\varepsilon$ is the scaling
factor. One can check that the classical action corresponding to the
Hamiltonian $H$ is invariant under this transformation. See that the
Hamiltonian $H$ transform as $H\to (1/\varepsilon^2)H$. The
Lagrangian $L$ associated with the system also transforms the same
way, $L\to(1/\varepsilon)L$. It is now obvious that the action,
$\mathcal A= \int dt L$, will be scale invariant under the
transformation $\bf{r}\to\varepsilon\bf{r}$ and
$t\to\varepsilon^2t$. In quantum mechanics, it has the following
consequences. Let $\phi$ is an eigen-state of the Hamiltonian $H$
with eigenvalue $E$, i.e., $H\phi= E\phi$, then
$\phi_\varepsilon=\phi(\varepsilon \bf r)$ will also be an
eigen-state of the same $H$ but with energy, $E/\varepsilon^2$. The
ground state therefore has no lower bound, implying that it does not
have any bound state. It is however known from some physical
problems, for example binding of electron in polar molecule
\cite{giri3}, the near horizon states of a black hole \cite{govinda}
and other \cite{giri,pulak1,pulak2,pulak3} that inverse square
potential can bind particles. The theoretical interpretation of this
binding can be obtained in terms of nontrivial quantization, which
can be obtained by von Neumann method of self-adjoint extensions.

However once the inverse square problem  is considered in a
non-commutative plane, it looses its scale symmetry property due to
the presence of dimensional parameter $\Theta$. To first order in
the parameter, $\Theta$, the potential $V= \alpha/r^2$ in
non-commutative plane becomes  more singular, but then it belongs to
an interesting class of interaction $V_\mu=\mbox{g}/r^\mu$, $\mu>2$
studied in \cite{mako}. The interesting feature of the potential
$V_\mu$ is that it possesses a localized state at the threshold of
energy $E=0$. The states which has zero eigenvalue is usually
considered as a transition point  from bound states to scattering
states. But due to the asymptotic  nature of the potential of the
type $V_\mu$ they can form bound states \cite{jamil}, even at $E=0$.
Apart from scale symmetry, inverse square problem has even larger
symmetry, formed by three generators: the Hamiltonian $H$, the
Dilatation generator $\mathcal D$ and the conformal generator $K$.
It is called the  $SO(2,1)$ algebra: $[\mathcal D,H]= -i\hbar H$,
$[\mathcal D,K]= i\hbar K$, $[H,K]= 2i\hbar \mathcal D$
\cite{wyb,alfaro}. We showed that with the introduction of
non-commutativity the $so(2,1)$ symmetry of the system is broken
explicitly and however in commutative limit  the exact $so(2,1)$
symmetry is restored.

In the present article we  extend our discussion of Ref.
\cite{pulak} further and find out a generic boundary condition for
the zero energy localized state. The article is organized in the
following fashion: First, we consider the  inverse square
interaction on a plane and discuss briefly how it changes when the
co-ordinates of the plane become non-commutative. Second, we
consider the non-commutative Hamiltonian obtained to first order in
non-commutativity parameter $\Theta$. The possible bound sate
spectrum is discussed in terms of generic boundary conditions.
Finally, we conclude with some discussion.

We now consider a particle, interacting with the potential
$V=\alpha/r^2$, on a non-commutative plane of the form
\begin{eqnarray}
\left[\hat{x_1},\hat{x_2}\right]= 2i\Theta,~
\left[\hat{p_1},\hat{p_2}\right]=0,~
\left[\hat{x_i},\hat{p_j}\right]=
i\hbar\delta_{ij}\,.\label{algebra1}
\end{eqnarray}
However, the commutative limit $\Theta\to 0$ takes it to the
standard algebra
\begin{eqnarray}
\left[x_1,x_2\right]= 0,~\left[p_1,p_2\right]= 0,~
\left[x_i,p_j\right]=i\hbar\delta_{ij}\,.\label{algebra2}
\end{eqnarray}
It is useful to get a representation of the non-commutative
coordinates $(\hat{x_i},\hat{p_i})$ in terms of the coordinates
$(x_i,p_i)$. We choose a representation
\begin{eqnarray}
\nonumber \hat{x_1}&=& x_1 -\Theta p_2,~\hat{x_2}=x_2 +\Theta p_1\,,\\
\hat{p_1} &= &p_1,~\hat{p_2}=p_2\,, \label{}
\end{eqnarray}
for our purpose, but other representations are also possible.
The Hamiltonian on
non-commutative plane
\begin{eqnarray}
H_{NC}= {\hat{p_1}}^2 + {\hat{p_2}}^2 +
\alpha/{\hat{r}}^2\,,\label{pot1}
\end{eqnarray}
to first order in non-commutative parameter $\Theta$  can be written
as
\begin{eqnarray}
H_{NC}= {p_1}^2 + {p_2}^2 + \alpha/{r}^2 +
2\alpha\Theta(x_1p_2-x_2p_1)/r^4\,.\label{pot1}
\end{eqnarray}
The presence of the potential $2\alpha\Theta(x_1p_2-x_2p_1)/r^4$
breaks the scale invariance. We solved the eigenvalue problem
\begin{eqnarray}
H_{NC}\psi_{NC} =E_{NC}\psi_{NC}\,,\label{eigen}
\end{eqnarray}
for $E_{NC}=0$ and found a bound state with angular momentum $m$ for
$\xi=\sqrt{\alpha +m^2} >1$ \cite{pulak}. For large values of the
non-commutative parameter, $\Theta$, it is also possible to get the
expectation values of the Hamiltonian. Since the zero energy
Schr\"odinger equation is exactly solvable it is possible to ask
what is the most general boundary condition in this case. To be
explicit, we consider an eigen-value problem of the form
\begin{eqnarray}
\widehat{H_{NC}}\psi_{NC}\equiv -\frac{r^4}{\alpha m}\left(p_1^2
+p_2^2 +
\alpha/r^2\right)\psi_{NC}=2\Theta\psi_{NC}\,,\label{eigen1}
\end{eqnarray}
Note that the dimensional parameter $2\Theta$ has been considered as
the eigenvalue for our problem. All square-integrable solutions for
different values of the parameter $\Theta$ correspond to the
$E_{NC}=0$ degenerate states. Even for complex values of the
parameter $\Theta$ if the solution $\psi_{NC}$ is square-integrable
then it corresponds to the bound state with $E_{NC}=0$. Since our
assumption in (\ref{algebra1}) is that the parameter $\Theta$ is
real, we will  restrict the parameter space to real line. It can be
done if we can ensure that $\widehat{H_{NC}}$ is self-adjoint. From
now onward  the symmetric operator $\widehat{H_{NC}}$ will be
investigated and a suitable boundary condition will be found out,
which will make the operator self-adjoint.
\begin{figure}
\includegraphics[width=0.45\textwidth, height=0.22\textheight]{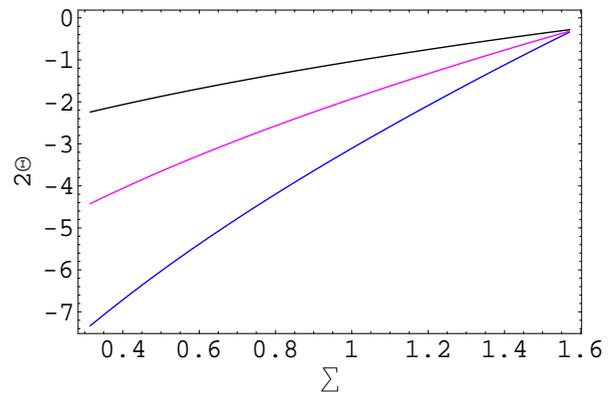}
\caption {(color online) A plot of the non-commutativity parameter $2\Theta$
  as a function of the self-adjoint extension parameter $\Sigma$ for
$m=1$. It correspons to the equation
(\ref{bound12}). The blue curve corresponds to $\alpha= -1/10$, the pink
curve corresponds to $\alpha=-1/6$ and the black curve corresponds to $\alpha=-1/3$.}
\end{figure}

Imposing a well defined boundary condition is important for getting
a physical solution. We in this article we exploit von Neumann's method
to analyze  $\widehat{H_{NC}}$.  So, before actually making any
symmetric extensions for the operator $\widehat{H_{NC}}$ a brief
discussion about the von Neumann's method is necessary here.
Consider  any symmetric operator, say, $\mathcal{B}$, which is for
the moment taken  to be unbounded. It is possible to define a domain
$D(\mathcal{B})$ under which the operator $\mathcal{B}$ is
symmetric. One can also obtain the adjoint operator,
$\mathcal{B}^*$, corresponding to the operator $\mathcal{B}$. From
the symmetric condition
$\int_0^\infty\phi^*(r)\mathcal{B}\chi(r)dr=\int_0^\infty
\left(\mathcal{B}^*\phi(r)\right)^*\chi(r)dr$, $\forall \chi(r)\in
D(\mathcal{B})$ we can obtain the domain, $D(\mathcal{B}^*)$. The
operator $\mathcal{B}$ would be self-adjoint if the two domains are
same, i.e., $D(\mathcal{B})= D(\mathcal{B}^*)$. In terms of the
deficiency indices $n_\pm$ \cite{reed} one can have alternative
definition of self-adjointness. The deficiency indices $n_\pm$ are
the dimension of the kernel $Ker(i\pm \mathcal{B}^*)$. If $n_\pm=0$,
then the operator $\mathcal{B}$ is essentially self-adjoint. If
$n_+=n_-=n\neq 0$, then $\mathcal{B}$ is not self-adjoint but admits
self-adjoint extensions. Self-adjoint extensions can be
characterized by $n^2$ parameters. Different values of the
parameters give rise to different physics. For, $n_+\neq n_-$, the
operator $\mathcal{B}$ does not have any self-adjoint extensions.

The operator, $\widehat{H_{NC}}$, we are analyzing in this work,
acts on the functions defined over the Hilbert space of
square-integrable functions with domain $\mathcal{L}^2[R^+, rdr]$.
Since the solution of the problem (\ref{eigen1}) has a similarity
with the inverse square problem \cite{giri}, $H\psi=E\psi$,  it
would be helpful to look at the short distance and asymptotic
behavior of both the solutions. One can check that  the solutions
have an inverse relation to each other of the form
\begin{eqnarray}
\nonumber \lim_{r\to 0}\psi_{NC} &\equiv & \lim_{r\to \infty}\psi\,,\\
\lim_{r\to \infty}\psi_{NC}&\equiv &\lim_{r\to 0}\psi\,.
\label{relation}
\end{eqnarray}
Due to this inverse behavior of the eigen-state we  impose a
nontrivial boundary condition for our problem at $r=\infty$. The
operator $\widehat{H_{NC}}$ is essentially self-adjoint for
$\xi^2\geq1$ and has been discussed in \cite{pulak}. Since any
system is defined by a Hamiltonian and its corresponding domain, in
our case $\widehat{H_{NC}}$ for $\xi^2\geq1$ acts over the domain
\begin{eqnarray}
\mathcal{D}_0=\{\psi\in \mathcal{L}^2(rdr),
\psi(\infty)=\psi'(\infty)=0\}\,. \label{domain}
\end{eqnarray}
Note the difference that the same condition (\ref{domain}) was
imposed for the inverse-square problem \cite{giri} but at $r \to 0$.
Let us now investigate the operator for the interval
$\xi\in(-1,1)$. In this region $\widehat{H_{NC}}$ is not essentially
self-adjoint and therefore we need to make self-adjoint extensions
of the original domain, so that the Hamiltonian becomes
self-adjoint. We discuss the case $\xi\neq 0$ first, and  then
consider the case $\xi=0$  separately. The deficiency indices are
$<1,1>$  for $\xi\in(-1,1)$. Since  the number of deficiency space
solutions are same for both types, there exist a  self-adjoint
extensions, characterized by a parameter, $\Sigma$. The domain under
which $\widehat{H_{NC}}$ would be self-adjoint is given by
\begin{eqnarray}
\mathcal {D}_\Sigma = \{\mathcal {D}_0 +\psi_+ + e^{i\Sigma}
\psi_-\}\,. \label{d6}
\end{eqnarray}
The explicit form of the deficiency space solutions $\psi_\pm$ are
given by
\begin{figure}
\includegraphics[width=0.45\textwidth, height=0.22\textheight]{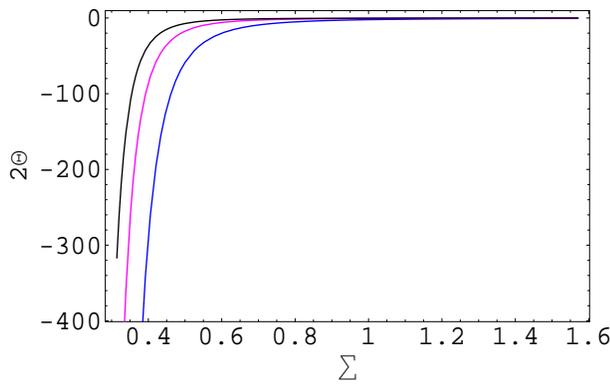}
\caption {(color online) A plot of equation (\ref{bound22}). Blue graph
  corresponds to $\alpha=-4$ and $m=2$. Pink graph
  corresponds to $\alpha=-9$ and $m=3$.  Black  graph
  corresponds to $\alpha=-16$ and $m=4$.}
\end{figure}
\begin{eqnarray}
\psi_+ &=& H_\xi\left(\frac{\sqrt{\alpha m}}{r}e^{-i\pi/4}\right)\,,\\
\psi_- &=& H_\xi\left(\frac{\sqrt{\alpha m}}{r}e^{+i\pi/4}\right)\,,
\end{eqnarray}
where $H_\xi$ is the modified Bessel function \cite{abr}. The
behavior of any function, belonging to the domain $\mathcal
{D}_\Sigma$, near   $r\to \infty$ can be found from the behavior of
$\psi_+ + e^{i\Sigma}\psi_-$ at asymptotic limit.  Because the
domain $\mathcal {D}_0$ goes to zero at $r \to \infty$, it does not contribute
to the domain at $r\to \infty$. The
asymptotic behavior of the domain  is of the form
\begin{eqnarray}
\lim_{r\to\infty}\left(\psi_+ + e^{i\Sigma}\psi_-\right)\simeq
\mathcal{A}_+\left(2r\right)^{-\xi} +
\mathcal{A}_-\left(2r\right)^{\xi}\,,\label{domain1}
\end{eqnarray}
where, $\mathcal{A}_\pm= -\frac{(\alpha m)^{\pm \xi/2}\pi
i}{\sin(\pi\xi)} \frac{\cos(\frac{\Sigma}{2}\pm
\frac{\pi\xi}{4})}{\Gamma(1 \pm\xi)}$. The solution of
(\ref{eigen1}) have to be matched with (\ref{domain1}) to get the
relation of the non-commutativity parameter $\Theta$ with the
self-adjoint extension parameter $\Sigma$. We see that there is
exactly one bound state with the non-commutativity, $2\Theta$, and
eigenfunction, $\psi_{NC}$, being of the form
\begin{eqnarray}
\label{bound12}2\Theta &=& \frac{1}{\alpha
m}\sqrt[\xi]{\frac{\cos\frac{1}{4}\left(2\Sigma+ \xi\pi\right)}
{\cos\frac{1}{4}\left(2\Sigma- \xi\pi\right)}}\,,\\ \psi_{NC} &=&
\exp(im\phi)H_\xi\left(\frac{\sqrt{-2\Theta \alpha m}}{r}\right)\,.
 \label{bound1}
\end{eqnarray}
In FIG. 1 the behavior of the parameter $2\Theta$ as a function of the
self-adjoint extension parameter $\Sigma$ has been shown for three different
values of the coupling constant $\alpha$ and for fixed value of the
angular momentum quantum number $m$. 
Now let us come to the case for $\xi = 0$, which  can be handled
similarly. The non-commutativity parameter corresponding to the
bound state and the corresponding eigen-state are given by
\begin{eqnarray}
\label{bound22}2\Theta &=& \frac{1}{\alpha m}{\exp}\left(\frac{\pi}{2}  {\cot}
\frac{\Sigma}{2}\right)\,, \\\psi_{NC} &=& \exp(im\phi)K_0\left(
\frac{\sqrt{-2\Theta \alpha m}}{ r}\right)\,.
\end{eqnarray}
respectively, where $K_0$ \cite{abr} is the modified Bessel
function. In FIG. 2 the parameter $2\Theta$ of (\ref{bound22}) has been
plotted as a function of the self-adjoint extension parameter $\Sigma$ for
three sets of values of the pair $\alpha$ and $m$.

Finally, to first order in non-commutativity, $\Theta$, the inverse
square problem has been discussed as a toy model to illustrate the
connection of the boundary conditions with the strength of
non-commutativity. The exact solvability of the $E_{NC}=0$
eigen-state has been exploited to get a generic boundary condition
by making a suitable self-adjoint extensions for the problem. We treated the
non-commutativity $\Theta$ as the eigen-value and obtained a generic boundary
conditions under which the specra is restricted to the subspace of real axis.

\end{document}